\documentclass[%
 reprint,
 amsmath,amssymb,
 aps,
 prb,
]{revtex4-2}

\usepackage{graphicx}
\usepackage{dcolumn}
\usepackage{bm}
\usepackage{braket}

\begin{document}


\title{Universal features of canonical phonon angular momentum without time-reversal symmetry}

\author{Hisayoshi Komiyama}
\author{Shuichi Murakami}
\affiliation{
Department of Physics, Tokyo Institute of Technology, 2-12-1 Ookayama, Meguro-ku, Tokyo 152-8551, Japan
}

\date{\today}

\begin{abstract}
It is known that phonons have angular momentum, and when the time-reversal symmetry (TRS) is broken, the total phonon angular momentum in the whole system becomes nonzero.
In this paper, we propose that as an angular momentum of phonons for a crystal without TRS, we need to consider the canonical angular momentum, as opposed to the kinetic angular momentum in previous works.
Next, we show that the angular momentum of phonons without TRS exhibits universal behaviors near the $\Gamma$ point.
We focus on in-plane oscillations in two-dimensional crystals as an example.
By breaking the TRS, one of the acoustic phonon branches at the $\Gamma$ point acquires a gap.
We show that the angular momentum of its acoustic phonon with a gap has a peak with the height $\pm \hbar$ regardless of the details of the system. 
From this, we find that this peak height changes discontinuously by changing the sign of the TRS-breaking parameter.
\end{abstract}

\maketitle

\section{Introduction}

Phonons are quasiparticles which carry heat in solids.
Many studies on phonons have been conducted on systems with time-reversal symmetry (TRS).
In recent years, the phonon Hall effect (PHE) has been observed experimentally \cite{strohm2005phenomenological}.
The PHE is a phenomenon in which under a magnetic field, a temperature gradient induces a heat flow in a direction perpendicular to both the temperature gradient and the magnetic field.
From such experiments, phonons in systems with broken TRS have attracted attention in recent years.
Furthermore, the PHE has been studied theoretically \cite{liu2017model, qin2012berry, zhang2010topological, saito2019berry, kagan2008anomalous, susstrunk2016classification} and experimentally \cite{strohm2005phenomenological, inyushkin2007phonon, sugii2017thermal, xu2018topological} from a topological point of view, similar to the electron Hall effect.

On the other hand, one can introduce a phonon angular momentum due to the vibration of the atoms inside the crystal \cite{zhang2014angular}.
The phonon angular momentum vanishes in systems with TRS in equilibrium.
In a system with TRS but without inversion symmetry, the phonon angular momentum becomes zero in the entire system, but the phonon angular momentum in each mode has a nonzero value.
In particular, the phonon angular momentum is nonzero at the valleys in the momentum space.
These phonons at the valleys are called chiral phonons \cite{zhang2015chiral} and have been observed experimentally \cite{zhu2018observation}.
Furthermore, in a system without inversion symmetry, the phonon angular momentum of the entire system can be generated by a temperature gradient \cite{hamada2018phonon}.
On the other hand, in a system without TRS, the entire system has a nonzero phonon angular momentum \cite{zhang2014angular}.
One can break the TRS for phonons by the Lorentz force \cite{holz1972phonons, kariyado2015manipulation}, the Coriolis force \cite{wang2015coriolis}, and spin-phonon interaction \cite{zhang2010topological, capellmann1991spin, capellmann1989microscopic}.
When these effects break the TRS and the phonon angular momentum of the entire system acquires a nonzero value, it may contribute to the Einstein–de Haas effect \cite{zhang2014angular, Einstein:1915:ENAb}.
Furthermore, methods for generating phonon angular momentum have been studied from various perspectives \cite{zhang2014angular, hamada2018phonon, hamada2020phonon, streib2021difference}.
In addition, various related subjects such as spin relaxation \cite{garanin2015angular, nakane2018angular, streib2018damping}, orbital magnetization of phonons \cite{juraschek2017dynamical, juraschek2019orbital, cheng2020large}, and conversion between magnons and phonons \cite{guerreiro2015magnon, holanda2018detecting} have also been studied.

As explained above, the phonon angular momentum for a system without TRS is important in understanding the Einstein–de Haas effect.
In this paper, we first formulate the angular momentum of phonons for a crystal without TRS.
Here, we point out that one can define two angular momenta, a canonical angular momentum and a kinetic angular momentum.
We propose that we need to consider the canonical one, as opposed to previous works, because the canonical one is conserved.
Next, we show that the angular momentum of acoustic phonons in systems without TRS exhibits universal behaviors near the $\Gamma$ point.
For this purpose, we consider in-plane oscillations in a two-dimensional crystal.
As an example, we calculate the phonon band structure and the phonon angular momentum of a kagome-lattice model without TRS by applying a magnetic field and Lorentz force.
By breaking TRS, one of the acoustic phonon branches acquires a gap at the $\Gamma$ point, while the other remains gapless.
In the kagome-lattice model, the phonon angular momentum has a peak equal to $\pm \hbar$ at the $\Gamma$ point, and this peak changes discontinuously between $\pm \hbar$ across the magnetic field $h=0$.
We show that these behaviors of the angular momentum of the acoustic phonon near the $\Gamma$ point are universal properties that do not depend on the details of the system.

This paper is organized as follows.
In Sec.~\ref{section:TRS-breaking phonons}, we review the eigenequation of the TRS-breaking phonons.
In Sec.~\ref{section:Phonon angular momentum without TRS}, we  formulate the canonical angular momentum of phonons for a crystal without TRS and discuss its difference from the kinetic angular momentum.
In Sec.~\ref{section:phonon angular momentum near the gamma point}, first, using a kagome-lattice model as an example, we explain that the angular momentum of the acoustic phonon with a gap has a peak with the height $\pm \hbar$ at the $\Gamma$ point, and this peak changes discontinuously between $\pm \hbar$ by changing the sign of the TRS-breaking parameter.
Using an effective Hamiltonian, we explain the universal property of acoustic phonons near the $\Gamma$ point when the TRS-breaking effect is small.
In Sec.~\ref{section:conclusion}, we summarize this paper.

\section{\label{section:TRS-breaking phonons} TRS-breaking phonons}

In this section, we review the eigenvalue problem for phonons when the TRS is broken, following Ref.~\cite{liu2017model, susstrunk2016classification}.
We begin with a Lagrangian for phonons in a crystal in the harmonic approximation:
\begin{align}
L_{0} = \frac{1}{2} \sum_{l} \bm{\dot{u}}_{l}^{\mathrm{T}} \bm{\dot{u}}_{l} - \frac{1}{2}  \sum_{l, l'} \bm{u}_{l}^{\mathrm{T}} K_{l, l'} \bm{u}_{l'}
\end{align}
where
$\bm{u}_{l} = (\vec{u}_{l,1}, \vec{u}_{l,2}, \cdots, \vec{u}_{l,n})^{\mathrm{T}}$,
$\vec{u}_{l,b}$ is a displacement vector of the $b$th atom in the $l$th unit cell multiplied by the square root of the mass of the atom, 
$n$ is the number of atoms in the unit cell,
and $K_{l, l'}$ is a mass-weighted force constant matrix.
From this Lagrangian, we get the eigenequation of the phonon: $D(\vec{k}) \bm{\epsilon}_{\vec{k}, \sigma} = \omega_{\vec{k}, \sigma}^{2} \bm{\epsilon}_{\vec{k}, \sigma}$, 
where
$D(\vec{k}) = \sum_{l'} K_{l, l'} e^{i (\vec{R}_{l'} - \vec{R}_{l}) \cdot \vec{k}}$ is the dynamical matrix,
$\omega_{\vec{k}, \sigma}$ is the eigenfrequency,
and $\bm{\epsilon}_{\vec{k}, \sigma}$ is the eigenstate of the eigenequation in the wave vector $\vec{k}$, specified by the mode index $\sigma = 1, 2, \cdots, N$.
Here $N$ is the dimension of the vector $\bm{u}_{l}$, and is given by $N = n d$, where $d$ is the dimension of the atomic displacement considered.
This eigenequation of the phonon assumes TRS \cite{maradudin1968symmetry}.

The TRS-breaking effect is treated by adding the term $L' = \sum_{l, l'} \dot{\bm{u}}_{l}^{\mathrm{T}} A_{l, l'} \bm{u}_{l'}$ to the Lagrangian $L_{0}$ \cite{liu2017model}.
According to Ref.~\cite{liu2017model}, $L' = \sum_{l, l'} \dot{\bm{u}}_{l}^{\mathrm{T}} A_{l, l'} \bm{u}_{l'}$ is the only harmonic term allowed when breaking the TRS for $L_{0}$, where $A_{l, l'}$ is a real matrix.
Furthermore, the symmetric part of $L'$ does not contribute to the motion because it can be written as the time derivative of $\frac{1}{2} \sum_{l, l'} \bm{u}_{l}^{\mathrm{T}} A^{\mathrm{S}}_{l, l'} \bm{u}_{l'}$ and contributes only a constant to the action $S = \int L dt$.
Therefore, when breaking the TRS of $L_{0}$, we consider only $L' = \sum_{l, l'} \dot{\bm{u}}_{l}^{\mathrm{T}} A_{l, l'} \bm{u}_{l'}$, where $A$ is a real antisymmetric matrix.
The physical origins of the TRS-breaking term for lattice vibration are the Lorentz force of charged ions \cite{holz1972phonons}, spin-phonon interaction in magnetic materials \cite{zhang2010topological}, and the Coriolis force with rotation \cite{wang2015coriolis,kariyado2015manipulation}.
In this paper, we consider that the Lorentz force breaks the TRS of charged ions in a lattice.
In this case, $\left( A_{l, l} \right)_{b \alpha, b \beta} = \frac{q_{b}}{2 m_{b}} \sum_{\gamma} \epsilon_{\alpha \beta \gamma} B_{\gamma}$, and the other elements are zero, 
where $m_{b}$ and $q_{b}$ are the mass and the charge of the $b$th atom; 
$\epsilon_{\alpha \beta \gamma}$ is the Levi-Civita symbol; 
$\alpha, \beta, \gamma$ run over $x$, $y$, $z$; 
and $B_{\gamma}$ is the magnetic field in the $\gamma$ direction.
From the Lagrangian $L = L_{0} + L'$,
the eigenequation without TRS becomes $D(\vec{k}) \bm{\epsilon}_{\vec{k}, \sigma} - 2 i \omega_{\vec{k}, \sigma} A \bm{\epsilon}_{\vec{k}, \sigma} = \omega_{\vec{k}, \sigma}^{2} \bm{\epsilon}_{\vec{k}, \sigma}$,
where $A$ is a real antisymmetric matrix.
This equation is not a generalized eigenproblem.
Therefore, we can make it a generalized eigenproblem by rewriting it as
\begin{align}
    \mathcal{H}(\vec{k}) \bm{\psi}_{\vec{k}, \sigma} &= \omega_{\vec{k}, \sigma} \bm{\psi}_{\vec{k}, \sigma}, \\
    \mathcal{H}(\vec{k}) &= \left(
        \begin{array}{cc}
          0 & i D(\vec{k})^{1/2} \\
          -i D(\vec{k})^{1/2} & -2iA
        \end{array}
      \right), \\
    \bm{\psi}_{\vec{k}, \sigma} &= \left(
        \begin{array}{c}
          \frac{i}{\sqrt{2} \omega_{\vec{k}, \sigma}} D(\vec{k})^{1/2} \bm{\epsilon}_{\vec{k}, \sigma} \\
          \frac{1}{\sqrt{2}} \bm{\epsilon}_{\vec{k}, \sigma}
        \end{array}
    \right).
\end{align}
This equation is called the \textit{Schr\"{o}dinger-like equation of phonons} because it is Hermitian.
We call $\mathcal{H}(\vec{k})$ the Hamiltonian in the following.

The dimension of $\mathcal{H}(\vec{k})$ is double the dimension $N$ of the dynamical matrix $D(\vec{k})$.
Since $\mathcal{H}(\vec{k})^{*} = -\mathcal{H}(-\vec{k})$, the eigenvalues $\omega_{\vec{k}, \sigma}$ and the eigenvectors $\bm{\psi}_{\vec{k}, \sigma}$ at the wavevector $\vec{k}$ can be labeled to satisfy $\omega_{\vec{k}, \sigma} = - \omega_{- \vec{k}, - \sigma}$ and $\bm{\psi}_{\vec{k}, \sigma}^{*} = \bm{\psi}_{- \vec{k}, - \sigma}$, where $\sigma$ is a band index $\sigma = -N, \cdots, -2, -1, 1, 2, \cdots, N$.
Therefore, there is one-to-one correspondence between the modes with negative frequencies and those with positive frequencies.
Because these two modes forming a pair represent one physical mode, we need to consider only the modes with positive frequencies in order to study their physical properties.
The normalization condition for the eigenstates is $\bm{\epsilon}_{\vec{k}, \sigma}^{\dag} \bm{\epsilon}_{\vec{k}, \sigma} + \frac{i}{\omega_{\vec{k}, \sigma}} \bm{\epsilon}_{\vec{k}, \sigma}^{\dag} A \bm{\epsilon}_{\vec{k}, \sigma} = 1$, which is rewritten as $\bm{\psi}_{\vec{k}, \sigma}^{\dag} \bm{\psi}_{\vec{k}, \sigma} = 1$.

\section{\label{section:Phonon angular momentum without TRS} Phonon angular momentum without TRS}

In this section, we first explain the angular momentum of phonons \cite{zhang2014angular}.
Next, we formulate the angular momentum of phonons without TRS.
The angular momentum of atoms in a crystal can be split into the mechanical angular momentum of the crystal as a rigid body and the angular momentum of the vibration of the atoms around their equilibrium position, and the latter is called phonon angular momentum.
We define the angular momentum of the vibration of the atoms in the crystal as 
\begin{align}
  \vec{J} = \sum_{l b} \vec{u}_{l b} \times \vec{p}_{l b}, 
\end{align}
where $\vec{p}_{l b}$ is a canonical momentum of the $b$th atom in the $l$th unit cell, divided by the square root of the mass of the atom.
The canonical momentum without TRS is $\bm{p}_{l} = \frac{\partial L}{\partial \bm{u}_{l}} = \bm{\dot{u}}_{l} + A \bm{u}_{l'}$, where $\bm{p}_{l} = (\vec{p}_{l,1}, \vec{p}_{l,2}, \cdots, \vec{p}_{l,n})^{\mathrm{T}}$.
In Ref.~\cite{zhang2014angular}, the angular momentum of phonons is defined by $\vec{J}^{\text{kin}} = \sum_{l b} \vec{u}_{l b} \times \dot{\vec{u}}_{l b}$, and precisely speaking, this should be called kinetic angular momentum when the TRS is broken.
Because the canonical angular momentum $\vec{J}$ is conservative but the kinetic one $\vec{J}^{\text{kin}}$ is not, we consider the canonical one in this paper.
We note that the matrix $A$ has a gauge degree of freedom, and the addition of any constant symmetric matrix to $A$ leaves the equation of motion invariant.
One may wonder if such a gauge degree of freedom exists also in the canonical angular momentum.
In Appendix~\ref{app:gauge invariant}, we discuss the gauge degree of freedom for a free charged particle in constant magnetic field $B$ along the $z$ axis.
We show that the canonical angular momentum along the $z$ axis is conserved only for a symmetric gauge with the vector potential $\vec{A} = \frac{1}{2} (- B y, B x, 0)$ and not conserved for other gauges.
Thus, in the discussion of the canonical angular momentum, we should fix the gauge to be a symmetric gauge.
In the present paper, we also adopt the symmetric gauge, which corresponds to the gauge with the matrix $A$ being antisymmetric: $\left( A_{l, l} \right)_{b \alpha, b \beta} = \frac{q_{b}}{2 m_{b}} \sum_{\gamma} \epsilon_{\alpha \beta \gamma} B_{\gamma}$

For simplicity, we focus on an in-plane oscillation in a two-dimensional crystal in the $xy$ plane.
As shown in Appendix~\ref{app:Calculation of the phonon angular momentum without TRS}, the canonical angular momentum of phonons without TRS of the whole crystal in the $z$ direction is expressed as
\begin{align}
  \label{total_PAM}
  J_{z} &= \sum_{\vec{k}, \sigma>0} l_{\vec{k}, \sigma} \left[ f(\omega_{\vec{k}, \sigma}) + \frac{1}{2} \right], \\
  \label{canonical angular momentum}
  l_{\vec{k}, \sigma} &= \hbar \bm{\epsilon}_{\vec{k}, \sigma}^{\dag} \left( M + \frac{i}{\omega_k} A M \right) \bm{\epsilon}_{\vec{k}, \sigma},
\end{align}
where 
$f(\omega_{\vec{k}, \sigma}) = 1 / (e^{\hbar \omega_{\vec{k}, \sigma} / k_{B} T}-1)$ is the Bose distribution function,
$T$ is the temperature,
$k_{B}$ is Boltzmann's constant,
$\hbar$ is Planck's constant,
$l_{\vec{k}, \sigma}$ is the angular momentum of a phonon of branch $\sigma$ at  wave vector $\vec{k}$,
$\bm{\epsilon}_{\vec{k}, \sigma}$ is a normalized eigenvector for the displacement vector,
and $M = I_{n \times n} \otimes \left(\begin{array}{cc}0 & -i \\ i & 0 \end{array} \right)$.
In contrast to Eq.~(\ref{canonical angular momentum}), in Ref.~\cite{zhang2014angular}, the angular momentum of a phonon is defined by $l_{\vec{k}, \sigma}^{\text{kin}} = \hbar \bm{\epsilon}_{\vec{k}, \sigma}^{\dag} M \bm{\epsilon}_{\vec{k}, \sigma}$, which is the kinetic angular momentum without TRS.
Therefore, the canonical angular momentum in Eq.~(\ref{canonical angular momentum}) has an additional term, $\frac{i \hbar}{\omega_k} \bm{\epsilon}_{\vec{k}, \sigma}^{\dag} A M  \bm{\epsilon}_{\vec{k}, \sigma}$, which makes the behavior of the angular momentum of acoustic phonons very different from previous studies when TRS is broken.

\section{Model calculation}

In this section, we calculate the phonon angular momentum of the kagome-lattice model without TRS as an example, and we discuss the temperature dependence of the phonon angular momentum.
To break the TRS, we calculate phonons in a model where atoms with electric charges form a kagome lattice and a magnetic field is applied in the out-of-plane direction.
The model is the same as the one studied in Ref.~\cite{zhang2014angular}, but the change in the definition of the angular momentum leads to different results from Ref.~\cite{zhang2014angular}.
At each of the three sublattices of the kagome-lattice model, we put atoms $A$, $B$, and $C$, depending on the sublattice.
Let the masses and the charges of atoms be $m_{i}$ and $q_{i}$ $(i = A, B, C)$, respectively.
When the TRS is broken by the Lorentz force under the magnetic field $h$ in the $z$ direction,
the Lagrangian acquires the term $L'$, with
\begin{align}
  A = \left(
    \begin{array}{cccccc}
      & - \frac{q_{A}}{2 m_{A}} h & & & & \\
      \frac{q_{A}}{2 m_{A}} h & & & & & \\
      & & & - \frac{q_{B}}{2 m_{B}} h & & \\
      & & \frac{q_{B}}{2 m_{B}} h & & & \\
      & & & & & -\frac{q_{C}}{2 m_{C}} h \\
      & & & & \frac{q_{C}}{2 m_{C}} h & \\
    \end{array}
  \right).
\end{align}

\begin{figure}
\includegraphics[clip,width=\linewidth]{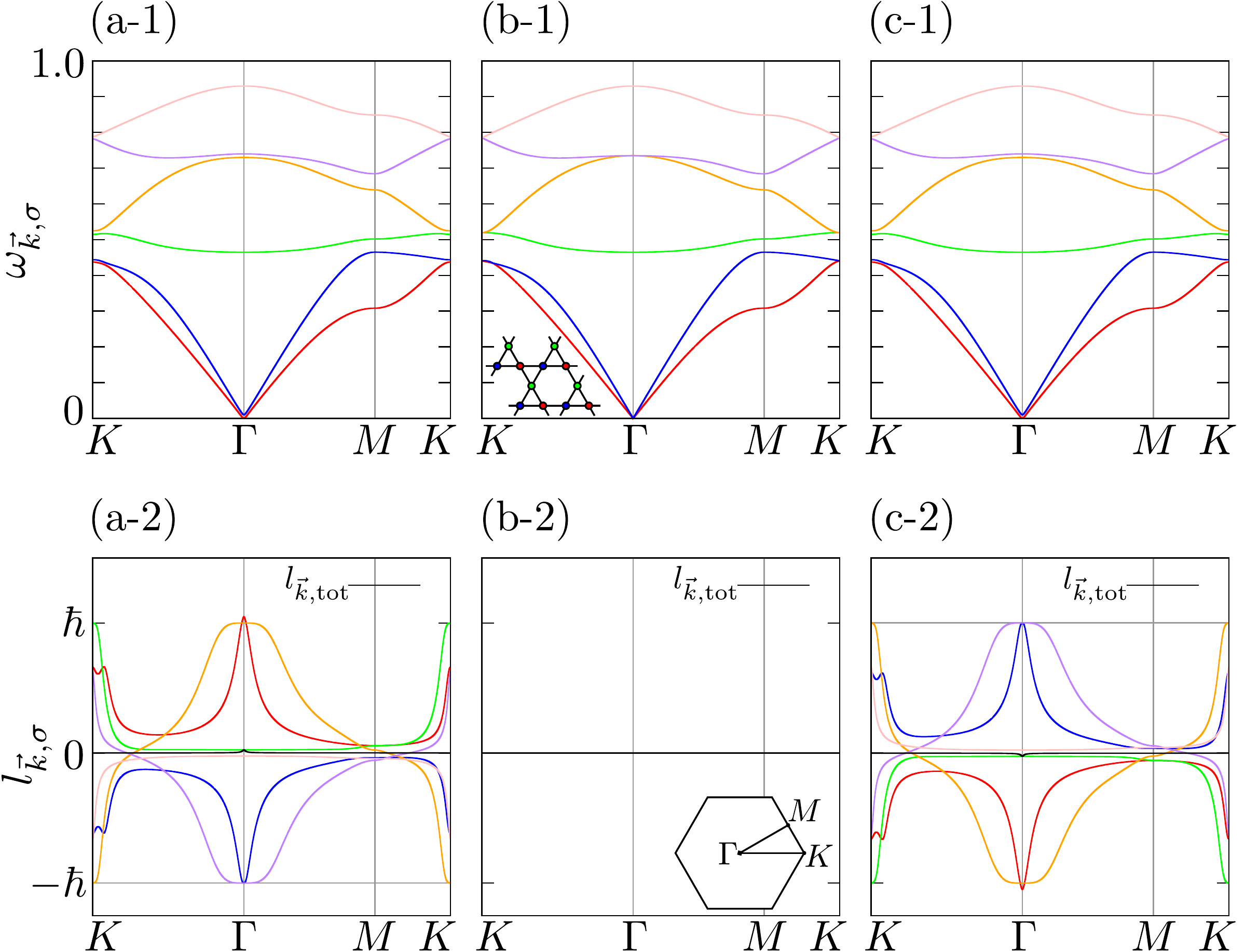}
\caption{\label{fig:honeycomb_h}
(a-1)-(c-1)The dispersion relation and (a-2)-(c-2) the angular momentum of the phonons in the kagome-lattice model when the magnetic field $h$ is varied.
The magnetic field $h$ is (a) $h=0.02$, (b) $h=0$, and (c) $h=-0.02$.
The colors of the curves represent the modes, and the black curves in (a-2), (b-2), and (c-2) represent the total angular momentum.
The inset in (b-1) shows the schematic pictures of the model, and that in (b-2) shows its Brillouin zone.
}
\end{figure}

We use the matrix $A$ to calculate its dispersion relation and angular momentum in a kagome lattice,
and the result is shown in Fig.~\ref{fig:honeycomb_h}.
In the calculation, we use the following values of the parameters:
the longitudinal spring constant $K_{L} = 0.144$,
the transverse one $K_{T} = K_{L} / 4$,
the lattice constant $a = 1$, and
the charge and mass of atoms $A, B, C$ are $m_{A} = m_{B} = m_{C} = 1, q_{A} = q_{B} = q_{C} = -1$.
The magnetic field $h$ in the $z$ direction is varied as $h =-0.02,\ 0,\ 0.02$.

\section{\label{section:phonon angular momentum near the gamma point}Phonon angular momentum at the $\Gamma$ point}

In this section, we show that the angular momentum of the acoustic phonon near the $\Gamma$ point shows a universal behavior that does not depend on the details of the system.
First, as an example, we calculate the phonon angular momentum of the kagome-lattice model without TRS, and we show that the angular momentum of the acoustic phonon shows a characteristic behavior.

As can be seen from Fig.~\ref{fig:honeycomb_h}, by adding a magnetic field, one of the two acoustic phonons with $\sigma = 2$ and $\vec{k} = 0$ acquires a gap at the $\Gamma$ point.
In addition, the angular momentum of this acoustic phonon is $\pm \hbar$.
When the magnetic field $h$ is changed from negative to positive, this angular momentum $l_{\Gamma, 2}$ changes discontinuously from $-\hbar$ to $\hbar$ across the magnetic field $h=0$.
Such a discontinuous change in the angular momentum is unexpected because an infinitesimal value of $h$ leads to a jump of the phonon angular momentum up to $\pm \hbar$.
As shown in Appendix~\ref{app:various systems}, this behavior of the angular momentum of acoustic phonons at the $\Gamma$ point holds in various systems in addition to the kagome-lattice model.
We note that similar calculations were performed for various modes in Ref.~\cite{zhang2014angular}, and the jump can be seen in Ref.~\cite{zhang2014angular}.
Nevertheless, the universality of the jump has not been noticed previously.
From now on, we show that this behavior is a universal property that does not depend on the details of the system.

We consider the behavior of acoustic phonons near the $\Gamma$ point when a TRS-breaking term is added to the phonon system.
The Schr\"{o}dinger-like equation of phonons with the TRS is
\begin{align}
  \mathcal{H}_{0} \bm{\psi}_{\sigma} &= \omega_{\sigma} \bm{\psi}_{\sigma}, \\
  \mathcal{H}_{0} &= \left(
      \begin{array}{cc}
        0 & i D(\Gamma)^{1/2} \\
        -i D(\Gamma)^{1/2} & 0
      \end{array}
    \right).
\end{align}
As mentioned above, for simplicity we focus on two-dimensional systems, with only the in-plane displacements.
Such a system has two acoustic phonon modes.
Accordingly, this eigenequation has four eigenvectors with $\omega_{\sigma}=0$.
Their eigenvectors $\bm{\phi}_{i} (i=1,2,3,4)$ are independent of the details of the system and are given by
\begin{align}
  \label{phi1234}
  \bm{\phi}_{1} = \left(
    \begin{array}{c}
      \bm{\epsilon}_{1} \\
      \hline
      \bm{0}
    \end{array}
  \right),
  \bm{\phi}_{2} = \left(
    \begin{array}{c}
      \bm{\epsilon}_{2} \\
      \hline
      \bm{0}
    \end{array}
  \right),
  \bm{\phi}_{3} = \left(
    \begin{array}{c}
      \bm{0}\\
      \hline
      \bm{\epsilon}_{1} \\
    \end{array}
  \right),
  \bm{\phi}_{4} = \left(
    \begin{array}{c}
      \bm{0}\\
      \hline
      \bm{\epsilon}_{2} \\
    \end{array}
  \right),
\end{align}

\begin{align}
  \bm{\epsilon}_{1} = C \left(
  \begin{array}{c}
    \sqrt{m_{1}} \\
    0 \\
    \sqrt{m_{2}} \\
    0 \\
    \vdots \\
    \sqrt{m_{n}} \\
    0
    \end{array}
  \right), \ 
  \bm{\epsilon}_{2} = C \left(
    \begin{array}{c}
      0 \\
      \sqrt{m_{1}} \\
      0 \\
      \sqrt{m_{2}} \\
      \vdots \\
      0 \\
      \sqrt{m_{n}}
    \end{array}
  \right), \ 
  C = \frac{1}{\sqrt{\sum_{\alpha} m_{\alpha}}},
\end{align}
where $m_{i}$ is the mass of the atom $(i = 1, \cdots, n)$ in the unit cell.
The forms of the eigenvectors are universal because they are Goldstone modes.
To consider the behavior of the acoustic phonons near the $\Gamma$ point when the TRS-breaking effect is small, we consider an effective $4 \times 4$ matrix for the Hamiltonian projected onto these four eigenvectors:
\begin{align}
  \label{eigenvalue matrix of acoustic phonon}
  \tilde{\mathcal{H}}(\vec{k}) = \left(
    \begin{array}{cccc}
      \bm{\phi}^{\dag}_{1} \mathcal{H}(\vec{k}) \bm{\phi}_{1} & \bm{\phi}^{\dag}_{1} \mathcal{H}(\vec{k}) \bm{\phi}_{2} & \bm{\phi}^{\dag}_{1} \mathcal{H}(\vec{k}) \bm{\phi}_{3} & \bm{\phi}^{\dag}_{1} \mathcal{H}(\vec{k}) \bm{\phi}_{4} \\
      \bm{\phi}^{\dag}_{2} \mathcal{H}(\vec{k}) \bm{\phi}_{1} & \bm{\phi}^{\dag}_{2} \mathcal{H}(\vec{k}) \bm{\phi}_{2} & \bm{\phi}^{\dag}_{2} \mathcal{H}(\vec{k}) \bm{\phi}_{3} & \bm{\phi}^{\dag}_{2} \mathcal{H}(\vec{k}) \bm{\phi}_{4} \\
      \bm{\phi}^{\dag}_{3} \mathcal{H}(\vec{k}) \bm{\phi}_{1} & \bm{\phi}^{\dag}_{3} \mathcal{H}(\vec{k}) \bm{\phi}_{2} & \bm{\phi}^{\dag}_{3} \mathcal{H}(\vec{k}) \bm{\phi}_{3} & \bm{\phi}^{\dag}_{3} \mathcal{H}(\vec{k}) \bm{\phi}_{4} \\
      \bm{\phi}^{\dag}_{4} \mathcal{H}(\vec{k}) \bm{\phi}_{1} & \bm{\phi}^{\dag}_{4} \mathcal{H}(\vec{k}) \bm{\phi}_{2} & \bm{\phi}^{\dag}_{4} \mathcal{H}(\vec{k}) \bm{\phi}_{3} & \bm{\phi}^{\dag}_{4} \mathcal{H}(\vec{k}) \bm{\phi}_{4}
    \end{array}
  \right).
\end{align}
We introduce $\vec{k} = k \vec{n}$, $k = |\vec{k}|$ and $\lambda = \bm{\epsilon}_{1}^{\dag} A_{\vec{0}} \bm{\epsilon}_{2}$ where $\lambda$ represents the magnitude of the TRS breaking.
To describe the phonons near the $\Gamma$ point when the TRS-breaking effect is small, we expand Eq.~(\ref{eigenvalue matrix of acoustic phonon}) up to linear order terms with respect to $\lambda, k$. 
We get
\begin{align}
  \label{eigenvalue matrix of acoustic phonon 2}
  \tilde{\mathcal{H}}(\vec{k}) \simeq \left(
    \begin{array}{cccc}
       & & i a_{\vec{n}} & i c_{\vec{n}}  \\
       & & i c_{\vec{n}} & i b_{\vec{n}}  \\
       -i a_{\vec{n}} & -i c_{\vec{n}} & & \\
       -i c_{\vec{n}} & -i b_{\vec{n}} & & 
    \end{array}
  \right) k + \left(
    \begin{array}{cccc}
       & & & \\
       & & & \\
       & & & -2i \\
       & & 2i & 
    \end{array}
  \right) \lambda,
\end{align}
where $a_{\vec{n}}$, $b_{\vec{n}}$, and $c_{\vec{n}}$ depend only on $\vec{n}$ and not on $k$.
One can directly show that $a_{\vec{n}},\ b_{\vec{n}}$ are real because $D(\vec{k})^{1/2}$ is a Hermitian matrix.
Furthermore, as we show in Appendix~\ref{app:property_of_effective_hamiltonian}, $c_{\vec{n}}$ is also real.
We note that in the calculation of Eq.~(\ref{eigenvalue matrix of acoustic phonon 2}) in Appendix~\ref{app:property_of_effective_hamiltonian}, the key step is how to calculate $D(\vec{k})^{1/2}$.
Due to a singularity of $k=0$, we need to separate the $\vec{k}$ dependence into $k$ and $\vec{n}$, by which we can safely take the square root of $D(\vec{k})$.
In addition, we define the eigenvalues of Eq.~(\ref{eigenvalue matrix of acoustic phonon 2}) as $\omega_{\vec{k}, 2}$, $\omega_{\vec{k}, 1}$, $\omega_{\vec{k}, -1}$, and $\omega_{\vec{k}, -2}$, starting with the larger eigenvalue.

\begin{figure}
  \includegraphics[clip,width=\linewidth]{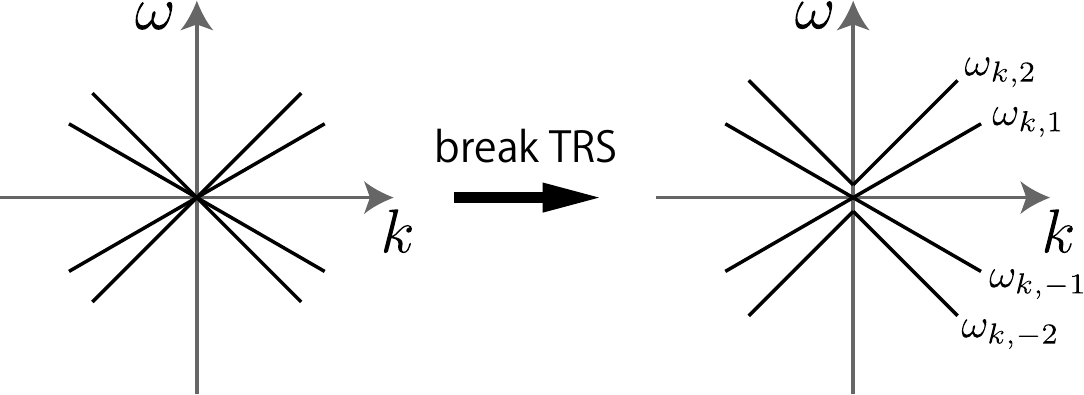}
  \caption{\label{fig:band_illust}
    Schematic picture of acoustic phonons when TRS is broken.
    When TRS is broken, one phonon has a positive frequency ($\omega_{\Gamma, 2}$), two phonons have zero frequencies ($\omega_{\Gamma, 1}$, $\omega_{\Gamma, -1}$), and one phonon has a negative frequency ($\omega_{\Gamma, -2}$) at the $\Gamma$ point.
  }
\end{figure}

Assuming $\lambda>0$, the eigenvector of the acoustic phonon with positive frequency $\omega_{\Gamma, 2} = 2 \lambda$ is $(0, 0, \frac{1}{\sqrt{2}}, \frac{i}{\sqrt{2}})^{\mathrm{T}}$.
Therefore, its eigenfunction is $\bm{\psi}_{\Gamma, 2} = \frac{1}{\sqrt{2}} \bm{\phi}_{3} + \frac{i}{\sqrt{2}} \bm{\phi}_{4} = \left( \begin{array}{c} \bm{0} \\ \frac{1}{\sqrt{2}} \bm{\epsilon}_{\Gamma, 2} \end{array} \right)$, where
\begin{align}
  \label{eps_plus}
  \bm{\epsilon}_{\Gamma, 2} = C \left(
    \begin{array}{c}
      \sqrt{m_{1}} \left(
        \begin{array}{c}
          1 \\
          i
        \end{array}
      \right) \\
      \sqrt{m_{2}} \left(
        \begin{array}{c}
          1 \\
          i
        \end{array}
      \right) \\
      \vdots \\
      \sqrt{m_{n}} \left(
        \begin{array}{c}
          1 \\
          i
        \end{array}
      \right)
    \end{array}
  \right).
\end{align}
Thus, this state represents the circular motions of all the atoms in the same phase \cite{watanabe2012unified}.
The angular momentum $l_{\Gamma, 2}$ of this eigenfunction $\bm{\psi}_{\Gamma, 2}$ is calculated to be $\hbar$.
Similarly, if $\lambda < 0$, the phonon mode $\omega_{\Gamma, 2} = -2 \lambda$ has the eigenfunction $\bm{\psi}_{\Gamma, 2} = \frac{1}{\sqrt{2}} \bm{\phi}_{3} - \frac{i}{\sqrt{2}} \bm{\phi}_{4}$, and its angular momentum is $l_{\Gamma, 2} = -\hbar$.
When $\lambda=0$, the acoustic phonons do not acquire a gap at the $\Gamma$ point.
Therefore, the angular momentum $l_{\Gamma, 2}$ of the acoustic phonon with a positive frequency at the $\Gamma$ point is determined by the sign of $\lambda$ as
\begin{align}
  l_{\Gamma, 2} \simeq \left\{ \begin{array}{ll}
     \hbar & (\lambda > 0), \\
     0 & (\lambda = 0), \\
     -\hbar & (\lambda < 0).
  \end{array} \right. .
\end{align}
This indicates that the angular momentum $l_{\Gamma, 2}$ changes discontinuously $\pm \hbar$ with respect to the parameter $\lambda$, which represents the magnitude of the TRS breaking.
Then this value is universal and independent of the details of the system.
On the other hand, the other two eigenvectors composed of $\bm{\phi}_{1}$ and $\bm{\phi}_{2}$ remain at $\omega = 0$ even without TRS.
This behavior of the frequencies is schematically shown in Fig.~\ref{fig:band_illust}.
The schematic picture in Fig.~\ref{fig:band_illust} agrees with the model calculation in Fig.~\ref{fig:honeycomb_h}.
Normally, one acoustic mode is gapped, while the other is gapless.

In the following, we physically interpret the value of the canonical angular momentum $l_{\Gamma, 2} = \hbar$ for the gapped mode at $\vec{k} = 0$ for the acoustic phonon with the positive frequency for $\lambda > 0$.
In the acoustic phonons at the $\Gamma$ point, relative positions of the atoms do not change.
Therefore, the springs between the atoms are not effective, and the motion is essentially the same as that of a free charged particle in a magnetic field.
In the present case, it is the cyclotron motion of a positive charge within the $xy$ plane, which is a simple and well-studied problem.
As presented in detail in Appendix~\ref{app:free charge}, in quantum mechanics, the motion is described in terms of two kinds of bosons, the $\hat{a}$ boson, with a positive frequency $\omega_C$ (cyclotron frequency) and a canonical angular momentum $+\hbar$, and the $\hat{b}$ boson, with zero frequency and a canonical angular momentum $-\hbar$.
These two bosons agree with the gapped mode and gapless mode in our calculations, respectively.
Indeed, in agreement with the $\hat{a}$ boson, the gapped mode has a canonical angular momentum $\hbar$ for $\lambda > 0$, which corresponds to the positively charged particle in Appendix C.
On the other hand, the gapless mode has a canonical angular momentum $- \frac{a_{\vec{n}}^2 + b_{\vec{n}}^2 + 2 c_{\vec{n}}^2} {2 a_{\vec{n}} b_{\vec{n}} - 2 c_{\vec{n}}^2} \hbar$, which is not equal to the one for the $\hat{b}$ boson.
This difference may be attributed to hybridization between the $\omega_{\vec{k}, 1}$ and $\omega_{\vec{k}, -1}$ branches at the $\Gamma$ point.
We also remark on the kinetic angular momentum.
We can calculate the kinetic angular momentum of phonons for the gapped and gapless modes to be $2 \hbar$ and $0$, respectively.
These values is again in complete agreement with results for the cyclotron motion of a free particle, as explained in Appendix~\ref{app:free charge}.

\section{\label{section:conclusion} Conclusion}

In this paper, we introduced a definition for the angular momentum of phonons without TRS modified from the one in Ref.~\cite{zhang2014angular}, 
and we showed that the angular momentum of acoustic phonons near the $\Gamma$ point without TRS shows universal behaviors that do not depend on the details of the system.
First, we pointed out that in the absence of TRS, apart from the kinetic angular momentum of phonons adopted in Ref.~\cite{zhang2014angular}, another angular momentum can be defined, called the canonical angular momentum of phonons.
Because the latter is conservative but the former is not, we considered the canonical angular momentum of phonons without TRS.
As an example, we calculated the band structure and angular momentum of phonons in a model of the kagome lattice under magnetic field, which breaks the TRS.
From this calculation, it was shown that the angular momentum of the acoustic phonon without TRS has a peak with a height $\hbar$ at the $\Gamma$ point.
The peak height changes sign when the sign of the magnetic field changes.
From these calculations, we predicted that the behavior of the angular momentum of the acoustic phonons near the $\Gamma$ point without TRS is universal, and we showed that that is, indeed, the case.

In order to prove the prediction for a general system, we introduced an effective Hamiltonian for phonons near the $\Gamma$ point with a small TRS-breaking effect.
Using this effective Hamiltonian, we showed that the acoustic phonon of the $\Gamma$ point with the breaking of TRS represents the circular motion of all the atoms in the same phase, and its angular momentum is $\pm \hbar$.
From this, we showed that this peak changes discontinuously by changing the sign of the TRS-breaking parameter.

\begin{acknowledgments}

This work was partly supported by JSPS KAKENHI Grant No.~JP18H03678.

\end{acknowledgments}

\appendix

\section{\label{app:Calculation of the phonon angular momentum without TRS} Calculation of the phonon angular momentum without TRS}

In this section, we derive Eqs.~(\ref{total_PAM}) and (\ref{canonical angular momentum}).
In the case of in-plane vibration of a two-dimensional system, the angular momentum of phonons is 
\begin{align}
  J_{z} = \sum_{l} \bm{u}_{l}^{\mathrm{T}} i M \bm{p}_{l}.
\end{align}
By using $\bm{p}_{l} = \bm{\dot{u}}_{l} + A \bm{u}_{l'}$ and the second quantization for $\bm{u}_{l}$,
\begin{align}
  \bm{u}_{l} = \sum_{\vec{k}, \sigma > 0} \sqrt{\frac{\hbar}{2 \omega_{\vec{k}, \sigma} N}} \bm{\epsilon}_{\vec{k}, \sigma} \hat{a}_{\vec{k}, \sigma} e^{i (\vec{k} \cdot \vec{R}_{l} - \omega_{\vec{k}, \sigma} t)} + \mathrm{H.c.} \ , 
\end{align}
we can obtain
\begin{align}
  J_{z} =& \sum_{\vec{k}, \sigma > 0} \hbar \bm{\epsilon}_{\vec{k}, \sigma}^{\dag} \left( M + \frac{i}{\omega_{\vec{k}, \sigma}} A M \right) \bm{\epsilon}_{\vec{k}, \sigma} \left( f(\omega_{\vec{k}, \sigma}) + \frac{1}{2} \right),
\end{align}
which is Eqs.~(\ref{total_PAM}) and (\ref{canonical angular momentum}) in the main text.

\section{\label{app:gauge invariant} Canonical angular momentum of a free charged particle}

In this appendix, we explain that the canonical angular momentum of a free charged particle moving in the $xy$ plane in a static magnetic field in the $z$ direction is conservative only in a symmetric gauge.
The Lagrangian of a free charged particle is 
\begin{align}
  \label{free particle lagrangian}
  L = \frac{m}{2} \dot{\vec{x}}^2 + \frac{q}{c} \dot{\vec{x}} \cdot \vec{A}(\vec{x}),
\end{align}
where $\vec{x} = (x, y)$ is the position vector of the particle,  
$m$ and $q$ are the mass and the charge of the particle, and 
$\vec{A}(\vec{x})$ is the vector potential.
Here the vector potential for the magnetic field $B$ in the $z$ direction is given by 
\begin{align}
  \vec{A}(\vec{x}) = \frac{1}{2} \left( \begin{array}{cc}
     & -B + \alpha \\
    B + \alpha & 
  \end{array} \right)
  \left( \begin{array}{c}
    x \\
    y
  \end{array} \right),
\end{align}
where $\alpha$ is a real constant representing the gauge degree of freedom.

By using Eq.~(\ref{free particle lagrangian}), the canonical momentum is $\vec{p} = \frac{\partial L}{\partial \dot{\vec{x}}}$, and the canonical angular momentum is $J = \vec{x} \times \vec{p}$.
It is written as $J = m (x \dot{y} - y \dot{x}) + \frac{q}{2 c} \left[ h (x^2+y^2) + \alpha (x^2-y^2) \right]$.
By using the equation of motion: $m \ddot{\vec{x}} = \frac{q}{c} \dot{\vec{x}} \times B$, the time derivative of the canonical angular momentum $J$ is 
\begin{align}
  \frac{d J}{d t} = \frac{q}{c} (x \dot{x} - y \dot{y}) \cdot \alpha.
\end{align}
Therefore, the canonical angular momentum in the static field is conservative only in $\alpha = 0$, that is, in a symmetric gauge.
Thus, while the vector potential $\vec{A}(\vec{x})$ allows a gauge degree of freedom, the canonical angular momentum should be defined in the symmetric gauge.
As in the case of a single charge, we define the canonical phonon angular momentum only in a symmetric gauge.

\section{\label{app:various systems} Model calculation of the phonon angular momentum}

\begin{figure*}
    \includegraphics[clip,width=\linewidth]{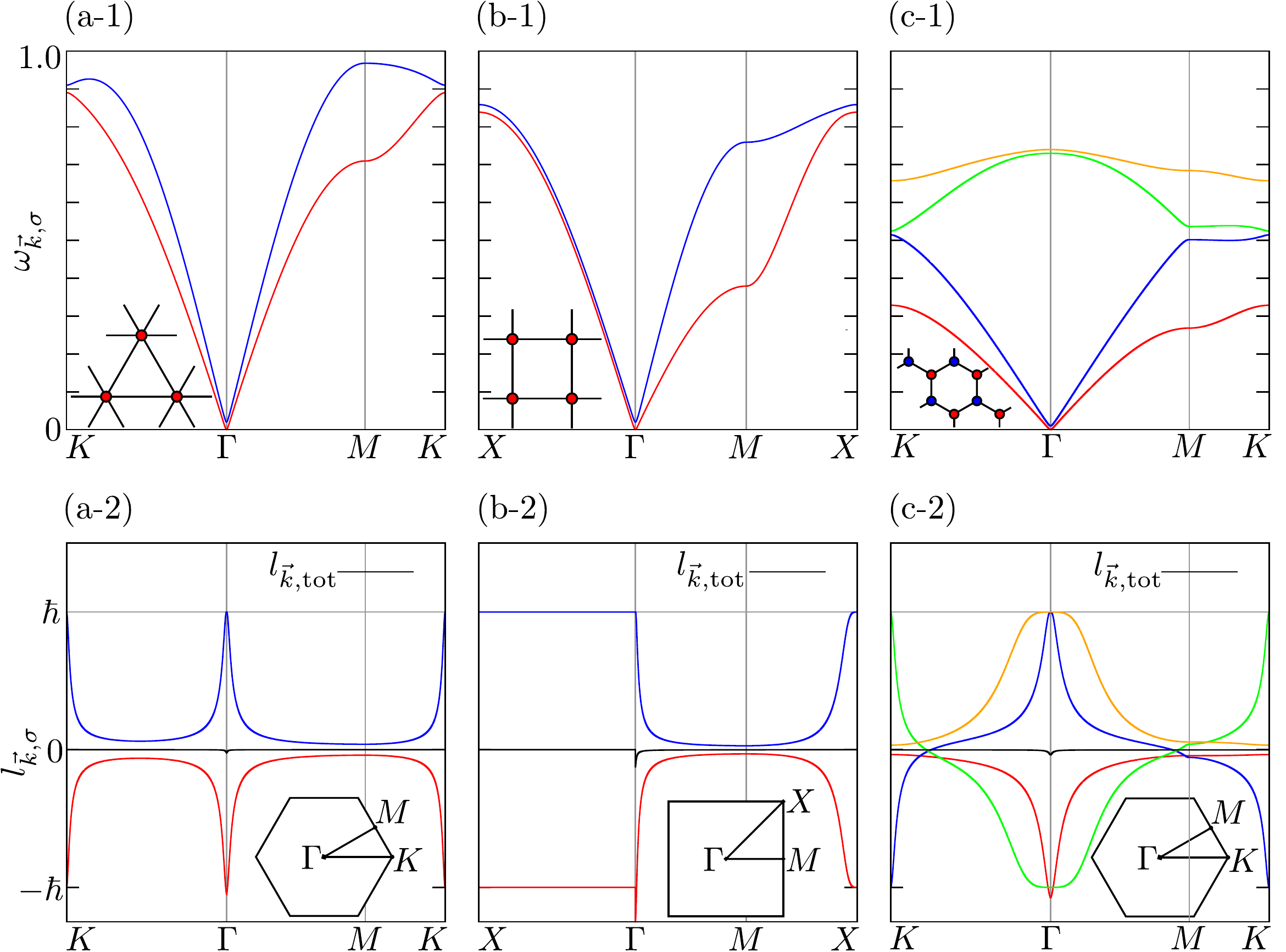}
    \caption{\label{fig:example}The dispersion relation and the angular momentum for (a) the triangle-lattice model, (b) the square-lattice model, and (c) The honeycomb-lattice model.
    We set the parameters as $h = 0.02$, $K_L = 0.144$, $K_T = K_L / 4$, $a = 1$ $m = 1$, and $q = -1$.
    The insets in (a-1), (b-1), and (c-1) show schematic pictures of the models, and those in (a-2), (b-2), and (c-2) show their Brillouin zones.}
\end{figure*}

In this appendix, as mentioned in Sec.~\ref{section:phonon angular momentum near the gamma point} we calculate various systems to confirm that the angular momentum of the acoustic phonon at the $\Gamma$ point, which acquires a gap through the TRS breaking, does not depend on the details of the system.
We calculate the band structure and the angular momentum for a triangular-lattice model, a square-lattice model, and a honeycomb-lattice model.
We note that a similar calculation was already performed in Ref.~\cite{zhang2014angular}.
The insets in Figs.~\ref{fig:example}(a-1)-\ref{fig:example}(c-1) show the triangular-lattice, square-lattice and honeycomb-lattice models, respectively.
In these models, one atom is located per each lattice site.
Let the mass and the charge of all atoms be $m$ and $q$, respectively.
We show the phonon frequency and angular momentum under the magnetic field $h$ in the direction perpendicular to the plane in Figs.~\ref{fig:example}(a-1) and \ref{fig:example}(a-2) for the triangular-lattice model, Figs.~\ref{fig:example}(b-1) and \ref{fig:example}(b-2) for the square-lattice model, and Figs.~\ref{fig:example}(c-1) and \ref{fig:example}(c-2) for the honeycomb-lattice model.
The values of the parameters used in the calculation in Fig.~\ref{fig:example} are the following:
the longitudinal spring constant $K_L = 0.144$,
the transverse one $K_T = K_L / 4$,
the lattice constant is $a = 1$,
the charge and mass of all atoms are $m=1, q=-1$, 
and the magnetic field $h = 0.02$.
As can be seen from Fig.~\ref{fig:example}, the dispersion relation and the angular momentum of the phonon differ depending on the model, but the dispersion relation and the angular momentum of the acoustic phonon at the $\Gamma$ point show the same behavior in all the models.
In particular, the angular momentum of the acoustic phonon at the $\Gamma$ point that appears after breaking the TRS has a peak with height $\hbar$ in all the models.

\section{\label{app:free charge} Free charged particle in a magnetic field}

In this appendix, we explain the cyclotron motion of a free charged particle in a uniform magnetic field.
In particular, we calculate its kinetic and canonical angular momenta in order to see the correspondence to the phonon angular momentum at the $\Gamma$ point discussed in the main text.
The Hamiltonian of a free charged particle in magnetic field $\vec{B} = (0, 0, B)$ is 
\begin{align}
  H &= \frac{1}{2 m} \vec{\Pi}^2, \\
  \vec{\Pi} &= \vec{p} + \frac{q}{c} \vec{A}(\vec{x}), 
\end{align}
where $m$ and $q$ are the mass and the charge of the particle,
$\vec{x}$ and $\vec{p}$ are the position vector and the canonical momentum of the particle, and
$\vec{A} (\vec{x}) = \frac{1}{2} \vec{B} \times \vec{x}$ is a vector potential.

We first introduce $(\xi, \eta)$ as 
\begin{align}
  (\xi, \eta) &= \frac{l^2}{\hbar} \left( \Pi_{y}, - \Pi_{x} \right), \\
  l &= \sqrt{\frac{c \hbar}{e B}}, 
\end{align}
which corresponds to the motion relative to the center of rotation.
This correspondence follows from the velocity $\vec{v} = (i / \hbar) \left[ H , \vec{x} \right] = \vec{\Pi} / m$ from the Heisenberg equation of motion, together with the relation $\vec{v} = \omega_{C} (-\eta, \xi)$ where $\omega_{C} = \frac{e B} {m c}$ is the cyclotron frequency.
We then introduce $(X, Y) \equiv (x - \xi, y - \eta)$ corresponding to the center of the rotation.

Then, the commutation relations among $\xi$, $\eta$, $X$, and $Y$ are
\begin{align}
  [\xi, \eta] =  -i l^2, &\ [X, Y] = i l^2, \\
  [\xi, X] = [\eta, Y] =& [\xi, X] = [\eta, Y] = 0.
\end{align}
From these relations we define two bosonic annihilation operators
\begin{align}
  \hat{a} &= \frac{-1}{\sqrt{2} l} (\eta + i \xi), \\
  \hat{b} &= \frac{1}{\sqrt{2} l} (X + i Y).
\end{align}
These two bosons commute: $[\hat{a}, \hat{b}] = [\hat{a}^{\dag}, \hat{b}] = 0$.
Therefore, by using the vacuum state $\ket{0}$, eigenstates can be written as $\ket{A, B} = (\hat{a}^{\dag})^A (\hat{b}^{\dag})^B \ket{0}$ with nonnegative integers $A$ and $B$.

Using these operators, we can define two number operators $\hat{N}_{a}$ and $\hat{N}_{b}$, for the two species of bosons:
\begin{align}
  \hat{N}_{a} &=  \hat{a}^{\dag} \hat{a} = \frac{1}{2 l^2} \left( \xi^2 + \eta^2 \right) - \frac{1}{2}, \\
  \hat{N}_{b} &= \hat{b}^{\dag} \hat{b} = \frac{1}{2 l^2} \left( X^2 + Y^2  \right) - \frac{1}{2}.
\end{align}
Then one can show 
\begin{align}
  H = \hbar \omega_C \left( \hat{N}_{a} + \frac{1}{2} \right), \\
  X^2 + Y^2 = 2 l^2 \left( \hat{N}_{b} + \frac{1}{2} \right).
\end{align}
Hence, the eigenstate $\ket{A, B}$ has an energy $E = \hbar \omega_{C} \left( A + \frac{1}{2} \right)$ and is also an eigenstate of the operator $X^2 + Y^2$ with an eigenvalue $2 l^2 \left( B + \frac{1}{2} \right)$.

By using the number operators, the canonical angular momentum can be written as 
\begin{align}
  L_{z} = \vec{x} \times \vec{p} = \hbar \left( \hat{N}_{a} - \hat{N}_{b} \right).
\end{align}
Therefore, the eigenstate $\ket{A, B}$ is an eigenstate of the canonical angular momentum with an eigenvalue $\hbar \left( A - B \right)$.
On the other hand, the kinetic angular momentum is defined as
\begin{align}
  L_z^{\text{kin}} = \vec{x} \times \vec{\Pi}.
\end{align}
Using the second quantized operators, we can express the kinetic angular momentum as
\begin{align}
  L_z^{\text{kin}} = 2 \hbar \left( \hat{N}_{a}+ \frac{1}{2} \right) + i \hbar \left( \hat{b} \hat{a} - \hat{b}^{\dag} \hat{a}^{\dag} \right).
\end{align}
Therefore, an expectation value of the kinetic angular momentum of the eigenstate $\ket{A, B}$ is $2 \hbar (A + \frac{1}{2})$.

Thus, to summarize, the $\hat{a}$ boson has an energy $\hbar \omega_{C}$, canonical angular momentum $\hbar$, and kinetic angular momentum $2 \hbar$, while the $\hat{b}$ boson has zero energy, canonical angular momentum $2 \hbar$, and zero kinetic angular momentum.
From this result, we can interpret the behavior of the angular momentum of the acoustic phonon at the $\Gamma$ point.
By breaking the TRS, one of the acoustic phonons acquires a nonzero frequency, while the other continues to have zero frequency at the $\Gamma$ point.
Thus, these phonons correspond to the $\hat{a}$ boson and the $\hat{b}$ boson, respectively.

\section{\label{app:property_of_effective_hamiltonian} Properties of the effective Hamiltonian~(\ref{eigenvalue matrix of acoustic phonon})}

In this appendix, we explain properties of the effective Hamiltonian of Eq.~(\ref{eigenvalue matrix of acoustic phonon}).
First, we explain the properties of the spring constant matrix $K_{l, l'}$ and the definition of the dynamical matrix $D(\vec{k})$.
Next, using these, we show that $a_{\vec{n}}$, $b_{\vec{n}}$, and $c_{\vec{n}}$ in Eq.~(\ref{eigenvalue matrix of acoustic phonon 2}) are real.
Furthermore, we explain how to calculate $a_{\vec{n}}$, $b_{\vec{n}}$, and $c_{\vec{n}}$ using the honeycomb-lattice model as an example.

\subsection{Properties of $D(\vec{k})$}

In this section, we explain the properties of the dynamical matrix $D(\vec{k})$, following Ref.~\cite{maradudin1968symmetry}.
When we expand the lattice potential $U$ in terms of the vector $\bm{u}_{l}$ around the equilibrium positions, which is the displacement multiplied by the square root of the mass of each atom in the unit lattice $l$ and extracted up to the second term, it becomes 
\begin{align}
    U \simeq U_{0}
    + \frac{1}{2} \sum_{l, l'} \bm{u}_{l}^{\mathrm{T}} K_{l, l'} \bm{u}_{l'},
\end{align}
where the first-order term becomes zero, and the coefficient in the second-order term is defined as
\begin{align}
  \left( K_{l, l'} \right)_{b \alpha, b' \beta} := \left. \frac{\partial^2 U}{\partial u_{l, b\alpha} \partial u_{l', b' \beta} } \right|_{\bm{u} = 0}.
\end{align}
Here, $u_{l, b\alpha}$ is the mass-weighted displacement of the $\alpha$ component of atom $b$ of the unit lattice $l$.

We explain the properties of $\left( K_{l, l'} \right)_{b \alpha, b' \beta} $.
First, it naturally follows from the definition that
\begin{align}
  \label{phi_1}
  \left( K_{l, l'} \right)_{b \alpha, b' \beta}  = \left( K_{l', l} \right)_{b' \beta, b \alpha} .
\end{align}
Next, the periodicity of the lattice yields
\begin{align}
  \label{phi_2}
  \left( K_{l, l'} \right)_{b \alpha, b' \beta} = \left( K_{0, l' - l} \right)_{b \alpha, b' \beta}.
\end{align}
Furthermore, the $\alpha$ component of the equation of motion of atom $b$ in the unit cell $l$ is given by
\begin{align}
  m_{l, b}\ddot{u}_{l, b\alpha}= - \sum_{l', b'\beta} \left( K_{l, l'} \right)_{b \alpha, b' \beta} u_{l', b' \beta}. 
\end{align}
Due to the translation symmetry, under uniform displacements of all the atoms, the force applied to atom $b$ in unit cell $l$ should be zero.
Therefore, we obtain
\begin{align}
  \label{phi_3}
  \sum_{l', b'} \left( K_{l, l'} \right)_{b \alpha, b' \beta} = 0 .
\end{align}
By using Eq. (\ref{phi_3}), we can rewrite the equation of motion,
\begin{align}
  m_{l,b}\ddot{u}_{l, b\alpha} =  \sum_{l', b' (\neq l, b)} \sum_{\beta} \left( K_{l, l'} \right)_{b \alpha, b' \beta} \left( u_{l, b\beta} - u_{l', b'\alpha} \right).
\end{align}
Therefore, the force applied by atom $b'$ of the unit cell $l'$ to atom $b$ of unit cell $l$ is
\begin{align}
  \sum_{\beta} \left( K_{l, l'} \right)_{b \alpha, b' \beta} \left( u_{l, b\beta} - u_{l', b'\alpha} \right) .
\end{align}
Similarly, the force applied by the atom $b$ of unit cell $l$ to atom $b'$ of unit cell $l'$ is
\begin{align}
  \sum_{\beta} \left( K_{l', l} \right)_{b' \alpha, b \beta} \left( u_{l', b' \beta} - u_{l, b\alpha} \right).
\end{align}
Since these two forces are in an action-reaction relationship, $\left( K_{l, l'} \right)_{b \alpha, b' \beta}$ satisfies
\begin{align}
  \label{phi_4}
  \left( K_{l, l'} \right)_{b \alpha, b' \beta} = \left( K_{l', l} \right)_{b' \alpha, b \beta}.
\end{align}

By using these properties of $\left( K_{l, l'} \right)_{b \alpha, b' \beta} $, the dynamical matrix $D(\vec{k})$ is defined as
\begin{align}
  \left( D(\vec{k}) \right)_{b\alpha, b'\beta} = \frac{1}{\sqrt{m_{b} m_{b'}}}\sum_{l'} \left( K_{l, l'} \right)_{b \alpha, b' \beta} e^{i \vec{k} \cdot \left( \vec{R}_{l'} - \vec{R}_{l} \right)}. 
\end{align}

\subsection{Expansion $D(\vec{k})^{1/2}$ with respect to $k = |\vec{k}|$}

Here, we will show that $a_{\vec{n}}$, $b_{\vec{n}}$, and $c_{\vec{n}}$  in the top right block in Eq.~(\ref{eigenvalue matrix of acoustic phonon 2}) are real.
First, this top right block of the matrix in Eq.~(\ref{eigenvalue matrix of acoustic phonon 2}) is written as
\begin{align}
  \label{def_abc}
  \left(
    \begin{array}{cc}
      \bm{\epsilon}_{1}^{\dag} D(\vec{k})^{1/2} \bm{\epsilon}_{1} & \bm{\epsilon}_{1}^{\dag} D(\vec{k})^{1/2} \bm{\epsilon}_{2}  \\
      \bm{\epsilon}_{2}^{\dag} D(\vec{k})^{1/2} \bm{\epsilon}_{1} & \bm{\epsilon}_{2}^{\dag} D(\vec{k})^{1/2} \bm{\epsilon}_{2}  
    \end{array}
  \right) \simeq
  \left(
    \begin{array}{cc}
      a_{\vec{n}} & c_{\vec{n}}  \\
      c_{\vec{n}}^{*} & b_{\vec{n}}  
    \end{array}
  \right) k
\end{align}
by extracting the terms up to linear order in $k$.
Here, because $D(\vec{k})$ is a positive-definite Hermitian matrix by definition, $D(\vec{k})^{1/2}$ is also a positive-definite Hermitian matrix.
Therefore, $a_{\vec{n}}$ and $b_{\vec{n}}$ are real and positive.
In the following, we show that $c_{\vec{n}}$ is also real.

Since $D(\vec{k})$ is analytic by definition, we expand the dynamical matrix $D(\vec{k})$ in terms of $\vec{k}$:
\begin{align}
  \label{D_veck}
  D(\vec{k}) \simeq D_{0} + \sum_{i} D^{(1)}_{i} k_{i} + \sum_{i, j} D^{(2)}_{ij} k_{i} k_{j},
\end{align}
where $D(\vec{k})$, $D_{0}$, $D^{(1)}_{i}$, and $D^{(2)}_{ij}$ are $2n \times 2n$ matrices.
Since $D(\vec{k})$ is a Hermitian matrix and preserves TRS, $D(\vec{k})^{\dag} = D(\vec{k}) = D(- \vec{k})^{*}$ holds.
Therefore, $D_{0}$ and $D^{(2)}_{ij}$ are real symmetric matrices, and $D^{(1)}_{i}$ are purely imaginary Hermitian matrices.
Then, in order to take its square root, we rewrite Eq.~(\ref{D_veck}) as
\begin{align}
  D(\vec{k}) \simeq D_{0} + D^{(1)}_{\vec{n}} k + D^{(2)}_{\vec{n}} k^{2},
\end{align}
where $\vec{k} = k \vec{n}$.
Then it follows that $D^{(2)}_{\vec{n}}$ is a real symmetric matrix and $D^{(1)}_{\vec{n}}$ is a purely imaginary Hermitian matrix.

We consider $U^{\dag} D(\vec{k}) U$ using the real orthogonal matrix $U$ that diagonalizes the real symmetric matrix $D_{0}$,
\begin{align}
  \label{UDU_simeq}
  U^{\dag} D(\vec{k}) U \simeq \Lambda^2 + \left( U^{\dag} D^{(1)}_{\vec{n}} U \right) k + \left( U^{\dag} D^{(2)}_{\vec{n}} U \right) k^{2},
\end{align}
where $\Lambda$ is a diagonal matrix defined as
\begin{align}
  \Lambda = \left(
      \begin{array}{cccc}
         \omega_{1} & & & \\
          & \omega_{2} & & \\
          & & \ddots & \\
          & & & \omega_{n}
      \end{array}
    \right),
\end{align}
with $\omega_{i} \ (i = 1,2, \cdots, 2n)$ being phonon frequencies at $\vec{k} = 0$. It follows that $\omega_{1} = \omega_{2} = 0$ because they represent acoustic phonons.

Since the first two column vectors of $U$ are eigenvectors $\bm{\epsilon}_1$, $\bm{\epsilon}_2$ of the acoustic phonons at the $\Gamma$ point, the $2 \times 2$ block on the top left of the left side of Eq.~(\ref{UDU_simeq}) is $\bm{\epsilon}_{\alpha}^{\dag} D(\vec{k}) \bm{\epsilon}_{\beta}(\alpha, \beta = 1, 2)$.
Using Eqs.~(\ref{phi_1}), (\ref{phi_2}), (\ref{phi_3}), and (\ref{phi_4}), we can calculate the component $\bm{\epsilon}_{\alpha}^{\dag} D(\vec{k}) \bm{\epsilon}_{\beta}$ $(\alpha, \beta = 1, 2)$ as
\begin{align}
    \label{D_sin}
    \bm{\epsilon}_{\alpha}^{\dag} D(\vec{k}) \bm{\epsilon}_{\beta} = \frac{-4}{\sum_{b} m_{b}} \sum_{bb'} \sum_{l>0} \left( K_{0, l} \right)_{b \alpha, b' \beta} \sin^{2}\left(\frac{\vec{k} \cdot \vec{R}_{l}}{2}\right). 
\end{align}
Therefore, in the $2 \times 2$ block on the top left of the left side of Eq.~(\ref{UDU_simeq}), the zeroth- and first-order terms of $k$ vanish.

Here, in Eq.~(\ref{UDU_simeq}) the $2n \times 2n$ matrices $\Lambda^2$, $U^{\dag} D^{(1)}_{\vec{n}} U$, and $U^{\dag} D^{(2)}_{\vec{n}} U$ are rewritten as
\begin{align}
  U^{\dag} D(\vec{k}) U \simeq \left( \begin{array}{c|c}
    \Lambda_{\text{ac}}^2 &  \\
    \hline
    & \Lambda_{\text{op}}^2 \\
  \end{array} \right) + \left( \begin{array}{c|c}
    A_{1} & C_{1} \\
    \hline
    C_{1}^{\dag} & B_{1}
  \end{array} \right) k + \left( \begin{array}{c|c}
    A_{2} & C_{2} \\
    \hline
    C_{2}^{\dag} & B_{2}
  \end{array} \right) k^{2},
\end{align}
where $\Lambda_{\text{ac}} = 0$; $A_{1}$ and $A_{2}$ are $2\times2$ matrices;
$B_{1}$, $B_{2}$, and $\Lambda_{op}^2$ are $(n-2)\times(n-2)$ matrices;
and $C_{1}$ and $C_{2}$ are $2\times(n-2)$ matrices.
Here $\Lambda_{\text{ac}}$ and $\Lambda_{\text{op}}$ are diagonal matrices with their diagonal elements given by frequencies of acoustic and optical branches at the $\Gamma$ point, respectively.
From Eq.~(\ref{D_sin}), we get $A_{1} = 0$.
Due to the properties of $D(\vec{k})$ mentioned in Eq.~(\ref{D_veck}) together with reality of $U$,  $A_{2}$, $B_{2}$, and $C_{2}$ are real matrices and $B_{1}$ and $C_{1}$ are purely imaginary matrices.

Based on this discussion, we can expand $U^{\dag} D(\vec{k})^{1/2} U$ up to the first order with respect to $k$ as follows:
\begin{align}
    \label{UDU}
    U^{\dag} D(\vec{k})^{1/2} U \simeq \left( \begin{array}{c|c}
      O &  \\
      \hline
      & \Lambda_{op} \\
    \end{array} \right) + \left( \begin{array}{c|c}
      X & Z \\
      \hline
      Z^{\dag} & Y
    \end{array} \right) k,
\end{align}
where $O$ is a $2\times2$ zero matrix.
We note that the top left $2\times2$ block of this matrix, $X k$, is equal to Eq.~(\ref{def_abc}).
Here we will show that $X$ is a real matrix.
By comparing both sides of $\left(U^{\dag} D(\vec{k})^{1/2} U \right)^2 = U^{\dag} D(\vec{k}) U$ we obtain
\begin{align}
  C_1 &= Z \Lambda_{\text{op}}, \\
  A_2 &= X^2 + Z Z^{\dag},
\end{align}
which yields
\begin{align}
  \label{X_ACL}
  X = \left( A_{2} - C_{1} \Lambda_{op}^{-2} C_{1}^{\dag} \right)^{1/2}.
\end{align}
One can directly show that $A_{2} - C_{1} \Lambda_{op}^{-2} C_{1}^{\dag}$ is a real symmetric matrix.
In addition, $X$ is a Hermitian matrix, meaning that its eigenvalues are real.
Therefore, the matrix $A_{2} - C_{1} \Lambda_{op}^{-2} C_{1}^{\dag} = X^2$ is a positive-semidefinite matrix, and it can be diagonalized by a real orthogonal matrix with nonnegative eigenvalues.
Therefore, we conclude that $X$ is a real symmetric matrix.
By comparing Eqs.~(\ref{def_abc}) and (\ref{UDU}) and noting that the first two column vectors of $U$ are $\bm{\epsilon}_1$ and $\bm{\epsilon}_2$, the matrix $X k$ is equal to Eq.~(\ref{def_abc}), which leads to the conclusion that $a_{\vec{n}}$, $b_{\vec{n}}$ and $c_{\vec{n}}$ are real.

\subsection{Calculation of $a_{\vec{n}}$, $b_{\vec{n}}$, and $c_{\vec{n}}$ in the honeycomb-lattice model}

We show how to calculate $a_{\vec{n}}$, $b_{\vec{n}}$, and $c_{\vec{n}}$ for the honeycomb-lattice model, as an example.
With reference to the Supplemental Material of Ref.~\cite{zhang2010topological}, the dynamical matrix $D(\vec{k})$ of the honeycomb-lattice model is 
\begin{align}
  D(\vec{k}) &= \left( \begin{array}{cc}
    K_1 + K_2 + K_3 &     - K_2 \\
        - K_2       & K_1 + K_2 + K_3
  \end{array} \right) \nonumber \\
  &+ \left( \begin{array}{cc}
     O & O \\
     - K_3 & O
  \end{array} \right) e^{i \vec{k} \cdot \vec{a}_1}
  + \left( \begin{array}{cc}
    O & - K_3 \\
    O & O
  \end{array} \right) e^{-i \vec{k} \cdot \vec{a}_1} \nonumber \\
  &+ \left( \begin{array}{cc}
    O & O \\
    - K_1 & O
  \end{array} \right) e^{i \vec{k} \cdot \vec{a}_2}
  + \left( \begin{array}{cc}
    O & - K_1 \\
    O & O
  \end{array} \right) e^{-i \vec{k} \cdot \vec{a}_2},
\end{align}
where 
$a_1=(a, 0)$ and $a_2=(a/2, \sqrt{3} a/2)$ are primitive vectors;
$K_1$, $K_2$, and $K_3$ are defined using $K_x = \left( \begin{array}{cc} K_L & \\  & K_T \end{array} \right)$ as $K_1 = U(\pi/2) K_x U(-\pi/2)$, $K_2 = U(\pi/6) K_x U(-\pi/6)$, $K_3 = U(-\pi/6) K_x U(\pi/6)$;
and $U(\theta) = \left( \begin{array}{cc} \cos \theta & - \sin \theta \\  \sin \theta & \cos \theta \end{array} \right) $ is a rotation matrix by an angle $\theta$ in two dimensions.

Next, we expand $D(\vec{k})$ with respect to $k$ as
\begin{align}
  \label{honeycomb_D_K}
  D(\vec{k}) \simeq \left( \begin{array}{cc}
     K & -K \\
    -K & K
  \end{array} \right) 
  &+ \left( \begin{array}{cc}
     O     & i K^{(1)}_{\vec{n}} \\
     -i K^{(1)}_{\vec{n}} & O
  \end{array} \right) k \nonumber \\
  &+ \left( \begin{array}{cc}
     O   & K^{(2)}_{\vec{n}} \\
     K^{(2)}_{\vec{n}} & O
  \end{array} \right) k^2,
\end{align}
where
$K = K_1 + K_2 + K_3 = \frac{3}{2} \left( \begin{array}{cc} K_{L} + K_{T} & 0 \\ 0 & K_{L} + K_{T} \end{array} \right)$,
$K^{(1)}_{\vec{n}} = (\vec{a}_1 \cdot \vec{n}) K_3 + (\vec{a}_2 \cdot \vec{n}) K_1$ and
$K^{(2)}_{\vec{n}} = \frac{(\vec{a}_1 \cdot \vec{n})^2}{2} K_3 + \frac{(\vec{a}_2 \cdot \vec{n})^2}{2} K_1$.
Since $K$ is a diagonal matrix, the orthogonal matrix $U$ that diagonalizes the first term in Eq.~(\ref{honeycomb_D_K}), i.e., the matrix $D_{0}$ in Eq.~(\ref{D_veck}), is
\begin{align}
  U = \frac{1}{\sqrt{2}}\left( \begin{array}{cc}
    I & I \\
    I & -I
 \end{array} \right),
\end{align}
where $I$ is the $2\times2$ identity matrix.
Therefore, we obtain
\begin{align}
  U^{\dag} D(\vec{k}) U \simeq \left( \begin{array}{cc}
    O & O \\
    O & 2K
  \end{array} \right) 
  &+ \left( \begin{array}{cc}
     O     & -i K^{(1)}_{\vec{n}} \\
     i K^{(1)}_{\vec{n}} & O
  \end{array} \right) k \nonumber \\
  &+ \left( \begin{array}{cc}
     K^{(2)}_{\vec{n}} & O \\
     O & -K^{(2)}_{\vec{n}}
  \end{array} \right) k^2.
\end{align}
Using Eq.~(\ref{X_ACL}), $ a_ {\vec {n}}, b_ {\vec {n}}, c_ {\vec {n}} $ can be calculated as
\begin{align}
  \label{honeycomb_abc}
  \left( \begin{array}{cc}
    a_{\vec{n}} & c_{\vec{n}} \\
    c_{\vec{n}} & b_{\vec{n}}
  \end{array} \right)
  = \left( K^{(2)}_{\vec{n}} - \frac{1}{2} K^{(1)}_{\vec{n}} K^{-1} K^{(1)}_{\vec{n}} \right)^{1/2},
\end{align}
where the matrix $K^{(2)}_{\vec{n}} - \frac{1}{2} K^{(1)}_{\vec{n}} K^{-1} K^{(1)}_{\vec{n}}$ is a real symmetric matrix.
We can actually express $K^{(2)}_{\vec{n}} - \frac{1}{2} K^{(1)}_{\vec{n}} K^{-1} K^{(1)}_{\vec{n}} = \left( \begin{array}{cc} \alpha_{\vec{n}} & \gamma_{\vec{n}} \\ \gamma_{\vec{n}} & \beta_{\vec{n}} \end{array} \right)$ as
\begin{align}
  \alpha_{\vec{n}} &= \frac{3 K_L^2 x^2 + K_T^2 (x-2y)^2 + 12 K_L K_T (x^2 -x y + y^2)} {24 (K_L + K_T)}, \\
  \beta_{\vec{n}} &= \frac{3 K_T^2 x^2 + K_L^2 (x-2y)^2 + 12 K_L K_T (x^2 -x y + y^2)} {24 (K_L + K_T)}, \\
  \gamma_{\vec{n}} &= \frac{(K_T - K_L) x (x - 2 y)}{8 \sqrt{3}},
\end{align}
where $x = \vec{a}_1 \cdot \vec{n}$ and $y = \vec{a}_2 \cdot \vec{n}$.
Here, the eigenvalues of the matrix $K^{(2)}_{\vec{n}} - \frac{1}{2} K^{(1)}_{\vec{n}} K^{-1} K^{(1)}_{\vec{n}}$ are $\frac{K_L^2 + 3 K_L K_T}{8 K_T + 8 K_L} a^2$ and $\frac{K_T^2 + 3 K_L K_T}{8 K_T + 8 K_L} a^2$, which are both positive.
Therefore, the matrix $K^{(2)}_{\vec{n}} - \frac{1}{2} K^{(1)}_{\vec{n}} K^{-1} K^{(1)}_{\vec{n}}$ is a positive-definite matrix.

Hence, we can calculate $a_{\vec{n}}$, $b_{\vec{n}}$, and $c_{\vec{n}}$ using $\alpha_{\vec{n}}$, $\beta_{\vec{n}}$, and $\gamma_{\vec{n}}$ as 
\begin{align}
\left( \begin{array}{cc}
  a_{\vec{n}} & c_{\vec{n}} \\
  c_{\vec{n}} & b_{\vec{n}}
\end{array} \right)
= \frac{1}{\sqrt{ \alpha_{\vec{n}} + \beta_{\vec{n}} + 2 \sqrt{\delta_{\vec{n}}}}} \left( \begin{array}{cc}
    \alpha_{\vec{n}} + \sqrt{\delta_{\vec{n}}} & \gamma_{\vec{n}} \\
    \gamma_{\vec{n}} & \beta_{\vec{n}} + \sqrt{\delta_{\vec{n}}}
\end{array} \right),
\end{align}
where $\delta_{\vec{n}} = \alpha_{\vec{n}} \beta_{\vec{n}} - \gamma_{\vec{n}}^2 > 0$.
Therefore, $a_{\vec{n}}$, $b_{\vec{n}}$, and $c_{\vec{n}}$ are real.

\bibliography{main}

\end{document}